\documentclass[twocolumn,preprintnumbers,secnumarabic,showpacs,amsmath,amssymb]{revtex4}
\usepackage{docs}

\usepackage{graphicx}
\usepackage{dcolumn}

\usepackage{bm}

\begin{document}
\title{Interior of a Schwarzschild black hole revisited}%


\author{Rosa Doran}%
\email{rdoran@cosmo.fis.fc.ul.pt} \affiliation{Centro de
Astronomia
e Astrof\'{\i}sica da Universidade de Lisboa,\\
Campo Grande, Ed. C8 1749-016 Lisboa, Portugal}

\author{Francisco S. N. Lobo}%
\email{flobo@cosmo.fis.fc.ul.pt} \affiliation{Centro de Astronomia
e Astrof\'{\i}sica da Universidade de Lisboa,\\
Campo Grande, Ed. C8 1749-016 Lisboa, Portugal}

\author{Paulo Crawford}%
\email{crawford@cosmo.fis.fc.ul.pt} \affiliation{Centro de
Astronomia
e Astrof\'{\i}sica da Universidade de Lisboa,\\
Campo Grande, Ed. C8 1749-016 Lisboa, Portugal}

\begin{abstract}

The Schwarzschild solution has played a fundamental conceptual
role in general relativity, and beyond, for instance, regarding
event horizons, spacetime singularities and aspects of quantum
field theory in curved spacetimes.
However, one still encounters the existence of misconceptions and
a certain ambiguity inherent in the Schwarzschild solution in the
literature. By taking into account the point of view of an
observer in the interior of the event horizon, one verifies that
new conceptual difficulties arise.
In this work, besides providing a very brief pedagogical review,
we further analyze the interior Schwarzschild black hole solution.
Firstly, by deducing the interior metric by considering
time-dependent metric coefficients, the interior region is
analyzed without the prejudices inherited from the exterior
geometry. We also pay close attention to several respective
cosmological interpretations, and briefly address some of the
difficulties associated to spacetime singularities. Secondly, we
deduce the conserved quantities of null and timelike geodesics,
and discuss several particular cases in some detail. Thirdly, we
examine the Eddington-Finkelstein and Kruskal coordinates directly
from the interior solution. In concluding, it is important to
emphasize that the interior structure of realistic black holes has
not been satisfactorily determined, and is still open to
considerable debate.

\end{abstract}


\maketitle

\section{Introduction}

The Schwarzschild solution has proved to play a fundamental
importance in conceptual discussions of general relativity, and
beyond, for instance, regarding event horizons, spacetime
singularities and aspects of quantum field theory in curved
spacetimes. It has also been important providing the first
insights regarding the phenomenon of gravitational collapse
\cite{OppSnyd} and inspired the construction of theoretical models
of relativistic stars \cite{relativistic-stars}.
Before the mid-1960s, the object now known as a black hole, was
referred to as a collapsed star \cite{Thorne-etal} or as a frozen
star \cite{Novikov}, and it was only in 1965 that marked an era of
intensive research into black hole physics. Relatively to the
issue of experimental tests of the Schwarzschild solution, the
exterior geometry has been extremely successful in explaining, for
instance, the precession of Mercury's perihelion, and the
phenomenon of the bending of light, where the exterior
Schwarzschild gravitational field acts as a gravitational lens.

Despite of its important role, one still encounters, in the
literature, the existence of misconceptions and a certain
ambiguity inherent in the Schwarzschild solution.
For instance, a problematic aspect is the presence of an event
horizon, which in the Schwarzschild black hole solution acts as a
one-way membrane, permitting future-directed null or timelike
curves to cross only from the exterior to the interior region. It
acts as a boundary of all events which, in principle, may be
observed by an exterior observer. It is believed that the
gravitational collapse of a compact body results in a singularity
hidden beyond an event horizon. If the singularity were visible to
the exterior region, one would have a {\it naked singularity},
which would open the realm for wild speculation. This led to
Penrose's {\it cosmic censorship conjecture} \cite{Penrose}, which
stipulates that all physically reasonable spacetimes are globally
hyperbolic, forbidding the existence of naked singularities, and
only allowing singularities to be hidden behind event horizons.
The cosmic censorship conjecture has been an active area of
research and the source of considerable controversy.
For the interior black hole solution, a remarkable change occurs
in the nature of spacetime, namely, the external spatial radial
and temporal coordinates exchange their character to temporal and
spatial coordinates, respectively. Thus, the interior solution
represents a non-static spacetime, as the metric coefficients are
now time-dependent. This also implies that a singularity occurs at
a spacelike hypersurface, $t=0$. Thus, no observer, interior or
exterior to the Schwarzschild radius, will be able to observe the
formation, or for that matter, the physical effects of the
singularity \cite{Narlikar}. These aspects show the existence of
inconsistencies and a certain ambiguity inherent in the
Schwarzschild solution.

Still relatively to the issue of the black hole event horizon, a
widespread misconception in the literature is that a test particle
approaches the Schwarzschild radius at the speed of light for all
observers, and not as a limiting process for a static observer
located at the event horizon given by the null hypersurface
$r\rightarrow 2M$, where $M$ is the black hole mass. We shall use
geometrized units, i.e., $G=c=1$, for notational convenience,
throughout this paper. If one accepts that a particle has the
speed of light with respect to a static observer, at $r=2M$, then
using the local special relativity velocity composition law, the
observer concludes that the particle has the speed of light with
respect to all observers, which is another way of saying that in
the frame of a photon all particles have speed $v=1$. Of course,
the frame of the photon is not a physical frame. Indeed, it should
be emphasized that an observer cannot remain at rest at $r=2M$, as
it implies an infinite acceleration to do so. Another erroneous
extrapolation of $v \rightarrow 1$ at $r=2M$, is that timelike
particles attain velocities superior to the speed of light in the
black hole region $0<r<2M$, and thus justifying the existence of
tachyons, or for that matter of particles with velocities superior
to the speed of light in the framework of general relativity
\cite{Gold-Sabb}. All things considered, it should be stressed
once again that one can only use static observers in the spacetime
region characterized by $r>2M$. And it was shown that the test
particle does indeed cross the event horizon with a velocity less
than the speed of light \cite{Janis} with respect to an
appropriate physical observer. An exact general expression was
further deduced, in terms of the constants of motion of both a
test particle and an observer moving along radially infalling
geodesics \cite{CrawTereno}, given by
\begin{equation}
v^2\big|_{r=2M}=1-\frac{4E_1^2 E_2^2}{\left(E_1^2+E_2^2 \right)^2}
\,
\end{equation}
where $E_1$ and $E_2$ are the constants of motion for the test
particle and the observer, respectively. This expression shows
that the value of the velocity at $r=2M$ is indeed less than $1$,
unless $E_1$ or $E_2$ are zero or infinite (see Ref.
\cite{CrawTereno} for details, and Ref. \cite{Bolos} for further
discussions).

Relatively to spacetime singularities, a key aspect is whether
they are a disaster for the theory, as they imply the breakdown of
predictability. Various definitions of singularities can be
encountered in the literature, ranging from specific invariants,
constructed from the curvature tensor and its covariant
derivatives, becoming infinite, to the notion of geodesic
incompleteness. Accordingly, one encounters several attitudes to
spacetime singularities \cite{Earman}: Firstly, singularities are
mere artifacts of unrealistic and idealized models where
infinities occur. Secondly, general relativity entails
singularities, according to the Penrose-Hawking theorems in the
context of geodesic incompleteness, and fails to accurately
describe nature. Thirdly, one may have a more optimistic
viewpoint, as expressed by Misner, who views the existence of
singularities, ``not as proof of our ignorance, but as a source
from which we can derive much valuable understanding of
cosmology"\cite{Misner}.

Much of the scepticism related to the concepts of event horizons
and spacetime singularities, outlined above, and others such as
the information paradox, has inspired new and fascinating ideas
\cite{Mazur,gravastars,Dymnikova,darkstars}, namely by replacing
the interior solution, and thus, doing away with the problems
related to these issues. In this context, it is interesting to
note the emergence of a new picture for an alternative final state
of gravitational collapse, where an interior compact object is
matched to an exterior Schwarzschild vacuum spacetime, at or near
where the event horizon is expected to form. These alternative
models do not possess a singularity at the origin and have no
event horizon, as its rigid surface is located at a radius
slightly greater than the Schwarzschild radius. In particular, the
gravastar ({\it grav}itational {\it va}cuum {\it star}) picture,
proposed by Mazur and Mottola \cite{Mazur}, has an effective phase
transition at/near where the event horizon is expected to form,
and the interior is replaced by a de Sitter condensate. It has
also been argued that there is no way of distinguishing a
Schwarzschild black hole from a gravastar from observational data
\cite{AKL}. However, a realistic model for the gravitational
collapse leading to compact interior objects is still lacking. We
also emphasize that, although interesting in themselves, the
solutions that do away with the interior singularity and the event
horizon \cite{Mazur,gravastars,Dymnikova,darkstars} sweep the
inherent conceptual difficulties of black holes under the rug, and
that the interior structure of realistic black holes has not been
satisfactorily determined, being still open to considerable
debate.

In this work, we shall review and analyze the interior
Schwarzschild solution in some detail. In the literature, the
interior geometry is considered as a continuity of the exterior
geometry. Therefore, for instance, infalling test particles are
considered ending up at a central singularity located at $r=0$.
However, the latter singularity is a spacelike hypersurface, and
the test particles are not directed towards a privileged point,
but directed along a temporal direction, in order to not violate
causality. Thus, if one looks at the interior geometry as a
continuation of the exterior static solution, one comes across
some extremely interesting conceptual difficulties, that question
the very concept of a black hole. In this work we shall address
some of these difficulties. We shall start by obtaining the metric
for the interior region without inferring directly to the
traditional Schwarzschild solution. With this geometry at hand,
one may analyze this interior region, without the prejudices
inherited from the exterior region. However, we emphasize that we
shall compare both regions where it is convenient to do so. We
find new interesting features and peculiarities not encountered in
the literature, and show that this scenario can be interpreted as
a cosmological solution.

This paper is outlined in the following manner: Firstly, in
Section \ref{Sec:interior}, we deduce the interior metric by
considering time-dependent metric fields, paying close attention
to several cosmological interpretations of this solution. We also
address some of the difficulties associated to spacetimes
singularities, and argue that it is perhaps possible for an
observer to know if he is inside or outside the Schwarzschild
radius, by examining an invariant of a covariant derivative of the
curvature tensor. Secondly, in Sections \ref{Sec:conserved} and
\ref{Sec:geodesics}, we deduce the conserved quantities of null
and timelike geodesics in some detail, discussing several
particular cases. Thirdly, in Sections \ref{Sec:EK-coordinates}
and \ref{Sec:Kruskal-coordinates}, we analyze the
Eddington-Finkelstein and Kruskal coordinates directly from the
interior solution. Finally, we conclude in Section
\ref{Sec:conclusion}.

\section{Interior spacetime}\label{Sec:interior}

\subsection{Spacetime metric}

We shall be interested in the spacetime metric given by
\begin{equation}
ds^2=-B(z,t)\,dt^2+A(z,t)dz^2+F(z,t) \,d\Omega^2 \,,
\label{metric}
\end{equation}
where $d\Omega^2=d\theta ^2+\sin ^2{\theta} \, d\phi ^2$ and
$(t,z,\theta,\phi)$ are the spacetime coordinates. Assume that
$F(z,t)=F(t)$, so that the line element with $t={\rm const}$ and
$z={\rm const}$ describes a $2-$sphere with an area given by
$A=4\pi F(t)$. In particular, we shall consider the specific case
of $F(t)=t^2$.

To set the nomenclature, note that the mathematical analysis and
the physical interpretation will be simplified using a set of
orthonormal basis vectors. These may be interpreted as the proper
reference frame of a set of observers who remain at rest in the
coordinate system $(t,z,\theta,\phi)$, with $(z,\theta,\phi)$
fixed. Denote the basis vectors in the coordinate system as ${\bf
e}_{t}$, ${\bf e}_{z}$, ${\bf e}_{\theta}$, and ${\bf e}_{\phi}$.
Then, using the following transformation, ${\bf
e}_{\hat{\alpha}}=\Lambda ^{\beta}_{\; \hat{\alpha}}\; {\bf
e}_{\beta}$, with
\begin{equation}
(\Lambda ^{\beta}_{\; \hat{\alpha}})=\left[
\begin{array}{cccc}
B^{-1/2}&0&0&0 \\
0&A^{-1/2}&0&0 \\
0&0&t^{-1}&0 \\
0&0&0&(t \sin \theta)^{-1}
\end{array}
\right] \,,
\end{equation}
one finds
\begin{eqnarray}
\left \{ \begin{array}{l}
{\bf e}_{\hat{t}}=B^{-1/2} \;{\bf e}_{t}\\
{\bf e}_{\hat{z}}=A^{-1/2}\; {\bf e}_{z}\\
{\bf e}_{\hat{\theta}}=t^{-1} \;{\bf e}_{\theta}\\
{\bf e}_{\hat{\phi}}=(t \sin \theta)^{-1}\; {\bf e}_{\phi}\,.
              \end{array} \right.
\end{eqnarray}
In this basis the metric components assume their Minkowskian form,
$g_{\hat{\alpha}\hat{\beta}}=\eta _{\hat{\alpha}\hat{\beta}}={\rm
diag} (-1,1,1,1)$.

The Einstein field equation, in an orthonormal reference frame, is
given by
\begin{equation}
G_{\hat{\mu}\hat{\nu}}=8\pi \,T_{\hat{\mu}\hat{\nu}},
\end{equation}
where $T_{\hat{\mu}\hat{\nu}}$ is the stress-energy tensor and
$G_{\hat{\mu}\hat{\nu}}$ is the Einstein tensor, given by
$G_{\hat{\mu}\hat{\nu}}=R_{\hat{\mu}\hat{\nu}}-\frac{1}{2}g_{\hat{\mu}\hat{\nu}}R$.
$R_{\hat{\mu}\hat{\nu}}$ is the Ricci tensor, which is defined as
a contraction of the Riemann (or curvature) tensor,
$R_{\hat{\mu}\hat{\nu}}=R^{\hat{\alpha}}_{\;\; \hat{\mu}
\hat{\alpha}\hat{\nu}}$, and $R$ is the scalar curvature defined
as a contraction of the Ricci tensor,
$R=R^{\hat{\alpha}}{}_{\hat{\alpha}}$.

The Einstein tensor, given  in the orthonormal reference frame,
$G_{\hat{\mu}\hat{\nu}}$, yields for the metric (\ref{metric}) the
following non-zero components
\begin{eqnarray}
G_{\hat{t}\hat{t}}&=&\frac{\dot{A}}{ABt}+\frac{1}{t^2}
        +\frac{1}{Bt^2} \label{Einsteintt}\,, \\
G_{\hat{z}\hat{z}}&=&\frac{\dot{B}}{t B^2}- \frac{1}{t^2}
        -\frac{1}{Bt^2}  \label{Einsteinzz}\,,  \\
G_{\hat{z}\hat{t}}&=&\frac{B'}{t B\sqrt{AB}} \label{Einsteinzt}\,,  \\
G_{\hat{\theta}\hat{\theta}}&=&-\frac{1}{2AB}\Bigg[\frac{\dot{A}}{t}
-\frac{A\dot{B}}{Bt}+\ddot{A}-B''
      \nonumber    \\
&&-\frac{\dot{A}^2}{2A}+\frac{A'B'}{2A}
+\frac{(B')^2}{2B}-\frac{\dot{A}\dot{B}}{2B} \Bigg]\,,
\label{Einsteintheta}\\
G_{\hat{\phi}\hat{\phi}}&=&G_{\hat{\theta}\hat{\theta}}\,,
\label{Einsteinphi}
\end{eqnarray}
where a prime denotes a derivative with respect to the coordinate
$z$, and the overdot a derivative with respect to the temporal
coordinate, $t$.

We shall consider a vacuum solution, i.e.
$G_{\hat{\mu}\hat{\nu}}=0$. From the addition of Eqs.
(\ref{Einsteintt}) and (\ref{Einsteinzz}), we verify
\begin{equation}
G_{\hat{t}\hat{t}}+G_{\hat{z}\hat{z}}=\frac{1}{tB} \left(
\frac{\dot{B}}{B} +\frac{\dot{A}}{A} \right)=0  \,,
\end{equation}
so that the solution $AB=C$, with $C=C(z)$ is obtained. It is
possible to absorb the function $C(z)$, defining a new spatial
coordinate $\bar{z}=\sqrt{C}\,z$, so that without a significant
loss of generality one may set $C=1$. One may conclude from this
analysis that $A=1/B$.

From Eq. (\ref{Einsteinzt}),
\begin{equation}
G_{\hat{z}\hat{t}}=\frac{B'}{t B\sqrt{AB}}=0   \,,
\end{equation}
we verify $B'=0$, so that $B=B(t)$, implying $A=A(t)$. Note that
$B=B(t)$ is related to the proper time $d\tau^2=B(t)\,dt^2$, so
that one may impose that $B(t)>0$.

Now, substituting the relationship $A=1/B$ into Eq.
(\ref{Einsteintt}), one finally deduces that
\begin{equation}
A(t)=B^{-1}(t)=\frac{C_1}{t}-1  \,,
\end{equation}
where $C_1$ is a constant of integration with time dimension. From
the condition $B(t)>0$, and consequently $A(t)>0$, this solution
is only valid for $t<C_1$.

Defining the constant of integration as $C_1=2\xi$ the metric
(\ref{metric}) finally takes the form
\begin{eqnarray}
ds^2&=&-\left(\frac{2\xi}{t}-1
\right)^{-1}\,dt^2+\left(\frac{2\xi}{t}-1 \right)dz^2
    \nonumber  \\
&&+t^2 \,(d\theta ^2 +\sin ^2{\theta} \, d\phi ^2)  \,.
\label{intmetric}
\end{eqnarray}
The constant $\xi$ may be determined from a direct confrontation
with the exterior Schwarzschild solution, and is given by $\xi=M$,
where $M$ is the black hole mass. It may take the physical
significance of a characteristic time for the existence of
universes in the interior Schwarzschild solution, as may be
inferred from the cosmological interpretation of the interior
metric, given by Eq. \ref{metric}, which we consider in the next
subsection.

\subsection{Cosmological interpretation}\label{sec:cosmo}

This interior solution illustrates a particularly strange, yet
physically meaningful picture of the universe within the event
horizon. Thus, we shall consider some interesting astrophysical
and cosmological interpretations of this solution. A quick glance
at the metric (\ref{intmetric}) is enough to convince one that
this also corresponds to an anisotropic and homogeneous
cosmological solution. In fact, considering the Kantowski-Sachs
\cite{KS} solution given by
\begin{equation}
ds^2=-d\bar{t}^{\,2}+A^2(\bar{t}\,)\,dz^2 +C^2(\bar{t}\,)
\,(d\theta ^2 +\sin ^2{\theta} \, d\phi ^2)  \,,
\label{Kantowski-Sachs}
\end{equation}
where $A(\bar{t}\,)$ and $C(\bar{t}\,)$ are the scale factors of
the geometry, one verifies that both metrics are identical, by
taking into account the following transformation
\begin{equation}
d\bar{t}^{\,2}=\left(\frac{2\xi}{t}-1 \right)^{-1}\,dt^2   \,.
\end{equation}

An alternative approach would be to consider a time-dependent
parameter $\xi=\xi(t)$ \cite{Vollick}, so that one could
generalize metric (\ref{intmetric}) to
\begin{eqnarray}
ds^2&=&-\left[\frac{2\xi(t)}{t}-1 \right]^{-1}\,
dt^2+\left[\frac{2\xi(t)}{t}-1 \right]dz^2
    \nonumber  \\
&&+t^2 \,(d\theta ^2 +\sin ^2{\theta} \, d\phi ^2)  \,.
\label{gen-intmetric}
\end{eqnarray}
The Einstein tensor given in an orthonormal reference frame has
the following non-zero components
\begin{eqnarray}
G_{\hat{t}\hat{t}}&=&\frac{2\dot{\xi}}{t^2}  \,,  \\
G_{\hat{z}\hat{z}}&=&-\frac{2\dot{\xi}}{t^2} \,,    \\
G_{\hat{\theta}\hat{\theta}}&=&G_{\hat{\phi}\hat{\phi}}=-\frac{\ddot{\xi}}{t}
\,,
\end{eqnarray}
where the overdot denotes a derivative with respect to the time
coordinate $t$, as before. Note that this solution implies
$\rho=-p_z$, where $\rho$ and $p_z$ are the energy density and the
pressure along the $z-$direction, much in the spirit of Refs.
\cite{gravastars,Dymnikova}. Note that the geometry
(\ref{gen-intmetric}) where $\xi$ is time-dependent is not a
solution of the vacuum Einstein equations. In addition, even
though $\rho=-p_z$, they are time-dependent, i.e., the
corresponding ``cosmological constant'' is not constant at all.

The above-mentioned case provides some very interesting
cosmological solutions, in rather different contexts, however,
they shall be presented elsewhere \cite{Doran}. Several
cosmological scenarios have also been proposed, in which a
universe emerges from the interior of a black hole (see, for
instance, Ref. \cite{Easson} and references therein). In the
present work, we shall only consider several interesting
interpretations of universes within the Schwarzschild radius,
relatively to the metric (\ref{intmetric}).

Consider the interior solution as measured by an observer at rest
relatively to the space coordinates, i.e., $dz=d\theta=d\phi=0$.
In this case, from the metric (\ref{intmetric}), we have
\begin{equation}
d\tau=\pm \frac{dt}{\sqrt{2\xi/t-1}}   \,.
    \label{int-observer}
\end{equation}
For the positive sign, we have the solution
\begin{equation}
\tau=-\sqrt{t(2\xi-t)} +
\xi\arctan\left[\frac{t-\xi}{\sqrt{t(2\xi-t)}}\right]
+\frac{\pi\xi}{2}\,,
    \label{prop-time}
\end{equation}
where the constant of integration has been chosen to provide
$\tau=0$ for $t=0$. Note that as $t=2\xi$, we have $\tau=\xi \pi$,
so that as coordinate time increases, the proper time as measured
by observers at rest also increase.

\begin{figure}[h]
\centering
  \includegraphics[width=3.2in]{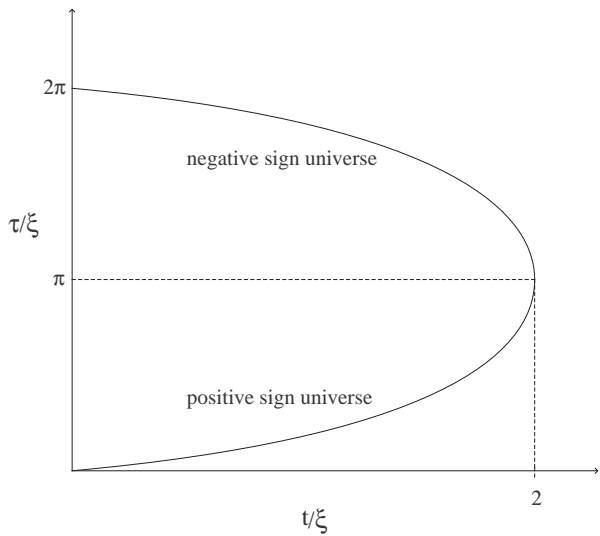}
  \caption{Plot of the proper time for observers at rest
  relatively to the $(z,\phi,\theta)$ coordinate
  system. The evolution of a universe begins at $t=0$,
  and proper time $\tau=0$. As coordinate time flows from $t=0$ to
  $t=2\xi$, proper time runs from $\tau=0$ to $\tau=\pi\xi$. This
  corresponds to the positive sign solution of Eq.
  (\ref{prop-time}). Now, allowing for the time coordinate to flow
  backwards from $t=2\xi$ to $t=0$, proper time as measured by
  observers at rest, inexorably runs forward from $\tau=\pi\xi$ to
  $\tau=2\pi\xi$. This corresponds to the negative sign solution of Eq.
  (\ref{prop-time2}). See the text for details.}
  \label{Fig:propertime}
\end{figure}

The evolution of this universe may be further explored
\cite{Brehme} by considering the negative sign of Eq.
(\ref{int-observer}), which yields the following solution
\begin{equation}
\tau=\sqrt{t(2\xi-t)} -
\xi\arctan\left[\frac{t-\xi}{\sqrt{t(2\xi-t)}}\right] +\frac{3\pi
\xi}{2}\,.
    \label{prop-time2}
\end{equation}
The constant of integration has been chosen to provide
$\tau=\xi\pi$ for $t=2\xi$. For this case the coordinate time
decreases from $t=2\xi$ to $t=0$, however, proper time increases
from $\tau=\pi \xi$ to $\tau=2\pi\xi$. This behavior is
represented in Fig. \ref{Fig:propertime}.

Taking into account the metric (\ref{intmetric}), we verify that
it possesses rotational invariance, as the spatial surfaces
corresponding to $z={\rm const}$, represent a $2-$sphere with an
area given by $A=4\pi t^2$. The proper distance between two
simultaneous events along a determined spatial direction, for
instance $d\phi=0$ and $\theta=\pi/2$, is given by
\begin{equation}
D_p=\int_{z_1}^{z_2}
\sqrt{\frac{2\xi}{t}-1}\;dz=\sqrt{\frac{2\xi}{t}-1}\;\Delta z \,.
\end{equation}
Note that a singularity occurs for $t=0$, as can also be verified
from the curvature tensor. The proper distance between two
particles at rest separated by a constant $\Delta z$, decreases
along the $z-$direction as coordinate time flows from $t=0$ to
$t=2\xi$, and increases as coordinate time flows backwards from
$t=2\xi$ to $t=0$. Despite the fact that one may not talk about an
asymptotic limit, for the interior solution, it is interesting to
note that the spacetime assumes an instantaneous Minkowski form,
for $t=\xi$, although the curvature does not become zero.

The proper distance between two simultaneous events along a
spatial trajectory with $dz=0$ and $\theta=\pi/2$, is given by
\begin{equation}
D_p=\int_{\phi_1}^{\phi_2}t\,d\phi=t\;\Delta \phi  \,,
\end{equation}
which increases as $t$ varies from $t=0$ to $t=2\xi$, and
decreases when the temporal coordinate runs backwards from
$t=2\xi$ to $t=0$.

Therefore, one may conclude by assuming the evolution of a
universe beginning at $t=0$, where a singularity occurs along the
$z-$direction, however, with no extension along the angular
direction, $\phi$. As coordinate time flows from $t=0$ to
$t=2\xi$, observers at rest move together, i.e., their proper
distance decreases to zero, and move apart along the angular
coordinate attaining a maximum at $t=2\xi$. Now, allowing for the
coordinate to flow backwards from $t=2\xi$ to $t=0$, proper time
as measured by observers, at rest relatively to the
$(z,\phi,\theta)$ coordinate system, inexorably runs forward from
$\tau=\pi\xi$ to $\tau=2\pi\xi$. For this case, observers move
apart along the $z-$direction and collapse along the angular
coordinate.

In this example, the difference of choosing an interior observer,
without the prejudices inherited from the exterior geometry, is
striking. While for the exterior observer, infalling particles
tend to a central singularity, from the interior point of view,
the proper distance along the $z-$direction increases, showing the
existence of a cigar-like singularity. We emphasize that the
latter occurs along a spacelike hypersurface. Another difference
worth mentioning is that the exterior observer considers a
spherically symmetric geometry, while the interior observer may
consider the geometry plane, as points for different $\phi$ are
parallel to one another (see Ref. \cite{Brehme} for details
regarding this issue).

\subsection{Singularities}

The Schwarzschild solution has played a fundamental role in
conceptual discussions of general relativity, in particular,
regarding spacetime singularities, as mentioned in the
Introduction. A key aspect of singularities in general relativity
is whether they are a disaster for the theory, as it implies the
breakdown of predictability. Attitudes in the literature range
from \cite{Earman}: singularities are mere artifacts of
unrealistic and idealized models; general relativity entails
singularities, but fails to accurately describe nature; and one
may view the existence of singularities ``not as proof of our
ignorance, but as a source from which we can derive much valuable
understanding of cosmology'', quoting Misner in the latter
attitude \cite{Misner}.

A way of detecting singularities is to find where the energy
density or the spacetime curvature become infinite and the usual
description of the spacetime breaks down. However, to be sure that
there is an essential singularity which cannot be transformed away
by a coordinate transformation, invariants are constructed from
the curvature tensor, such as $R$, $R_{\mu\nu}R^{\mu\nu}$,
$R^{\mu\nu\alpha\beta} R_{\mu\nu\alpha\beta}$, and from certain
covariant derivatives of the curvature tensor. For instance, in
the Schwarzschild spacetime there is an essential curvature
singularity at $r=0$ in the sense that along any non-spacelike
trajectory falling into the singularity, as $r\rightarrow 0$, the
so-called Kretschman scalar tends to infinity, i.e.,
$K=R^{\mu\nu\alpha\beta} R_{\mu\nu\alpha\beta} \rightarrow
\infty$, as shall be shown below. In this case, however, all
future directed non-spacelike geodesics which enter the event
horizon at $r=2M$ must fall into this curvature singularity within
a finite value of the affine parameter. So, all such curves are
future geodesically incomplete.
For the black hole region, given by the metric (\ref{intmetric}),
the scalar Kretschmann polynomial, $K$, is given by
\begin{equation}
R^{\mu\nu\alpha\beta} R_{\mu\nu\alpha\beta} =\frac{48\xi^2}{t^6}
\,,
\end{equation}
showing that a curvature singularity occurs at $t=0$.

It is remarkable that a change of sign occurs in the following
scalar \cite{KLA}, as an observer traverses the event horizon
\begin{equation}
R^{\mu\nu\alpha\beta;\gamma} R_{\mu\nu\alpha\beta;\gamma}
=-720\frac{(2\xi-t)\xi^2}{t^9} \,.
\end{equation}
Note that the invariant is zero on the horizon $t=2\xi$. It is
perhaps possible that this invariant is devoid of a fundamental
significance. However, it is generally known that using the
curvature tensor and some of its covariant derivatives, the
analysis gives a complete description of the geometry, and are
directly measurable. Since these quantities are coordinate
invariant, the problems associated with a specific choice of the
coordinate system vanish. This argument may be used in favor of
separating the interior from the exterior region.

\subsection{Tidal forces}

The tidal acceleration felt by an observer at rest is given by
\begin{equation}
\Delta
a^{\hat{\mu}}=-R^{\hat{\mu}}_{\;\;\hat{\nu}\hat{\alpha}\hat{\beta}}
\,U^{\hat{\nu}}\eta^{\hat{\alpha}}U^{\hat{\beta}}    \,,
\end{equation}
where $U^{\hat{\mu}}=\delta^{\hat{\mu}}_{\;\;\hat{0}}$ is the
observer's four velocity and $\eta^{\hat{\alpha}}$ is the
separation between two arbitrary parts of his body. Note that
$\eta^{\hat{\alpha}}$ is purely spatial in the observer's
reference frame, as $U^{\hat{\mu}}\eta_{\hat{\mu}}=0$, implying
$\eta^{\hat{t}}=0$.
$R^{\hat{\mu}}_{\;\;\hat{\nu}\hat{\alpha}\hat{\beta}}$ is the
Riemann tensor, given in the orthonormal reference frame, and has
the following non-zero components
\begin{eqnarray}
R_{\hat{z}\hat{\theta}\hat{z}\hat{\theta}}&=&
R_{\hat{z}\hat{\phi}\hat{z}\hat{\phi}}=-\frac{\xi}{t^3} \,, \\
R_{\hat{z}\hat{t}\hat{z}\hat{t}}&=&
-R_{\hat{\theta}\hat{\phi}\hat{\theta}\hat{\phi}}=-\frac{2\xi}{t^3}   \,,  \\
R_{\hat{\theta}\hat{t}\hat{\theta}\hat{t}}&=&
R_{\hat{\phi}\hat{t}\hat{\phi}\hat{t}}=\frac{\xi}{t^3}  \,.
\end{eqnarray}
Taking into account the antisymmetric nature of
$R^{\hat{\mu}}_{\;\;\hat{\nu}\hat{\alpha}\hat{\beta}}$ in its
first two indices, we verify that $\Delta a^{\hat{\mu}}$ is purely
spatial with the components
\begin{equation}
\Delta a^{\hat{i}}=-R^{\hat{i}}{}_{\hat{t}\hat{j}\hat{t}}
\,\eta^{\hat{j}}=-R_{\hat{i}\hat{t}\hat{j}\hat{t}}
\,\eta^{\hat{j}}. \label{spatial}
\end{equation}
Finally, using the components of the Riemann tensor, the tidal
acceleration has the following components
\begin{eqnarray}
\Delta a^{\hat{z}}&=&\frac{2\xi}{t^3}\,\eta^{\hat{z}} \,,  \\
\Delta a^{\hat{\theta}}&=&-\frac{\xi}{t^3}\,\eta^{\hat{\theta}}
\,,   \\
\Delta a^{\hat{\phi}}&=&-\frac{\xi}{t^3}\,\eta^{\hat{\phi}} \,.
\end{eqnarray}
Note a stretching along the $z-$direction, and a contraction along
the orthogonal directions. These stretchings and contractions are
now time-dependent, contrary to their counterparts in the exterior
region, and as $t\rightarrow 0$, the tidal forces diverge.

\section{Conserved quantities}\label{Sec:conserved}

Consider the Euler-Lagrange equations
\begin{equation}
\frac{d}{d\lambda}\left(\frac{\partial L}{\partial
\dot{x}^\mu}\right)-\frac{\partial L}{\partial x^\mu}=0 \,,
     \label{Euler-Lagrange}
\end{equation}
where the overdot here represents a derivative with respect to the
affine parameter defined along the geodesic, which has the
physical interpretation of a proper time for timelike geodesics.
Consider the following Lagrangian
\begin{equation}
L(x^\mu,\dot{x}^\mu)=\frac{1}{2}g_{\mu\nu}\dot{x}^\mu \dot{x}^\nu
\,.
\end{equation}
If the metric tensor does not depend on a determined coordinate,
$x^\mu$, one obtains an extremely important result. For this case,
Eq. (\ref{Euler-Lagrange}) reduces to
\begin{equation}
\frac{d}{d\lambda}\left(\frac{\partial L}{\partial
\dot{x}^\mu}\right)=0    \,.
     \label{Euler-Lagrange2}
\end{equation}
This implies that the quantity given by
\begin{equation}
p_\mu=\frac{\partial L}{\partial
\dot{x}^\mu}=g_{\mu\nu}\,\dot{x}^\nu  \,,
     \label{conserv-quant}
\end{equation}
is constant along any geodesic. Using the Lagrangian nomenclature,
one denotes $x^\mu$ a cyclic coordinate, and $p_\mu$ the
respective conjugate momentum. The existence of cyclic coordinates
allows one to obtain integrals of the geodesic equation, and
provides certain quantities that are conserved along the movement
of the particle.

Applying the above analysis to the line element (\ref{intmetric}),
one verifies that the metric tensor is independent of the
coordinates $z$ and $\phi$, so that the conserved quantities are
given by
\begin{eqnarray}
P_{\phi}&=&g_{\phi\phi}\,\dot{\phi}=t^2\dot{\phi}=Q  \,,  \label{P}  \\
P_z&=&g_{zz}\,\dot{z}= \left(\frac{2\xi}{t}-1 \right)\dot{z}
\label{Pz} \,.
\end{eqnarray}
$Q$ may be interpreted as the angular momentum per unit mass, and
$P_z$ possesses the dimensions of a velocity. As $P_z$ may take
any real value, we shall consider it as a mere conserved quantity,
without any physical significance.

The line element (\ref{intmetric}) may be rewritten in terms of
the constants defined above, for the particular case of
$\theta=\pi/2$, in the following manner
\begin{equation}
\dot{t}^2= P_z^2+\left(\frac{Q^2}{t^2}-k
\right)\,\left(\frac{2\xi}{t}-1 \right)   \,, \label{line1}
\end{equation}
where $k=0$ is defined for null geodesics, and $k=-1$ for timelike
geodesics.

For timelike geodesics, $k=-1$, the conserved quantities $P_z$ and
$Q$ may also be determined from the initial conditions. For this
purpose it will prove useful to provide an intrinsic definition of
velocity, which we shall include next for self-completeness.

Consider the four-velocity, $U^\mu$, tangent to the worldline of
an observer, and a four-dimensional spacetime, $\Sigma$,
orthogonal to $U^\mu$. Define the operator
\begin{equation}
h^\mu{}_{\nu}=g^\mu{}_{\nu}+U^{\mu}U_{\nu}  \,,
\end{equation}
which has the property of projecting any four-vector on the
tangent space of the hypersurface, $\Sigma$, so that
$h^\mu{}_{\nu}\,U^\nu=0$. Thus, one may express the metric tensor
in the following form
\begin{eqnarray}
ds^2&=&g_{\mu\nu}\,dx^\mu\,dx^\nu
      \nonumber     \\
&=&-\left(U_{\mu}\,dx^\mu \right)^2+h_{\mu\nu}\,dx^\mu\,dx^\nu
   \nonumber     \\
&=&-d\tau_*^{2}+dl^2    \,.
\end{eqnarray}
The quantity $d\tau_{*}=-U_{\mu}\,dx^\mu$ is the projection of the
displacement of a particle, $dx^\mu$, along the velocity of the
observer, so that the particle has a displacement of
$dl^2=h_{\mu\nu}\,dx^\mu\,dx^\nu$, along $\Sigma$. Thus, the
velocity may then be defined as
\begin{equation}
V^2=\left(\frac{dl}{d\tau_*}\right)^2
=\frac{h_{\mu\nu}\,dx^\mu\,dx^\nu}{\left(U_{\mu}\,dx^\mu
\right)^2}   \,.
        \label{velocity}
\end{equation}

Now, consider that the observer is at rest in the $x^\mu$
coordinates, so that his/her four-velocity is given by
$U^{\mu}=(U^t,0,0,0)$, with $U^t=(-g_{tt})^{-1/2}$. Thus, we have
\begin{equation}
h_{\mu\nu}=g_{\mu\nu}-\frac{g_{\mu t}g_{\nu t}}{g_{tt}}  \,,
\end{equation}
and
\begin{equation}
d\tau_*^2=-\frac{\left(g_{\mu t}\,dt\right)^2}{g_{tt}} \,,
\end{equation}
so that Eq. (\ref{velocity}) may be finally written as
\begin{eqnarray}
V^2=\frac{\left(g_{\mu t}g_{\nu
t}-g_{tt}g_{\mu\nu}\right)\,dx^\mu\,dx^\nu} {\left(g_{\mu
t}\,dx^\mu\,\right)^2} \,.
     \label{velocity2}
\end{eqnarray}
This result is identical to the one obtained by Landau and
Lifschitz \cite{Landau}.

Now, using the metric (\ref{intmetric}) and considering
$\theta=\pi/2$, Eq. (\ref{velocity2}) takes the form
\begin{equation}
V^2=-\frac{g_{zz}}{g_{tt}}\left(\frac{dz}{dt}\right)^2
-\frac{g_{\phi\phi}}{g_{tt}}\left(\frac{d\phi}{dt}\right)^2  \,.
\end{equation}
and finally using Eq. (\ref{line1}), we have
\begin{equation}
P_z^2=g_{zz}\left(\frac{V^2}{1-V^2}-\frac{Q^2}{t^2}\right) \,.
   \label{gen-geod}
\end{equation}

\begin{figure}[h]
\centering
  \includegraphics[width=3.0in]{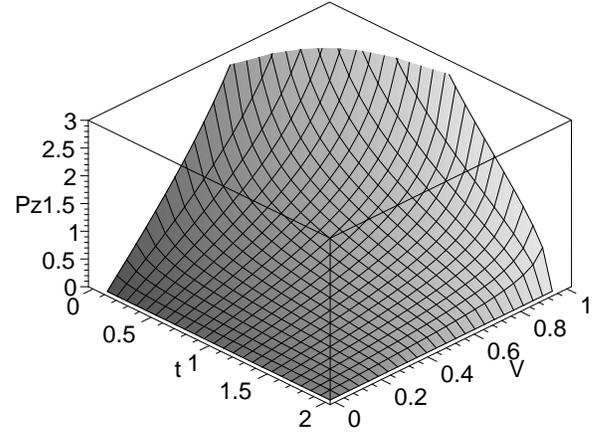}
  \caption{Qualitative behavior of the positive values of $P_z$ in the parameter
  space $t$ and $V$, where we have considered $\xi=1$. See the text for details.}
  \label{Fig:Pz-along-z}
\end{figure}

Considering the particular case of $d\theta=d\phi=0$, i.e., fo a
test particle that moves along the $z-$direction Eq.
(\ref{gen-geod}) takes the form
\begin{equation}
P_z^2=\left(\frac{2\xi}{t}-1\right)\left(\frac{V^2}{1-V^2}\right)
\,.
      \label{Pz(V)}
\end{equation}
The qualitative behavior for the positive values of $P_z$, in the
parameter space of $t$ and $V$, is represented in Fig.
\ref{Fig:Pz-along-z}. Note that $P_z$ may take arbitrarily large
values as $V\rightarrow 1$ or as $t\rightarrow 0$.

For a point particle with an initial velocity $V_0<1$ and initial
time $t=T$, then $P_z$ is given by
$P_z^2=(2\xi/T-1)V_0^2/(1-V_0^2)$. Note that if the test particle
is at rest, $V_0=0$, at an instant $T\neq 2\xi$, then it will
always remain at rest as $P_z=0$. If the particle came in from the
exterior region, it possesses a conserved quantity $E$ along its
geodesic. Despite the fact that after the crossing of the event
horizon its character changes into a constant with the dimensions
of a velocity, its numerical value is conserved, i.e., $E=P_z$.
The constant may assume different positive values depending on its
initial conditions. However, as reflected by Eq. (\ref{Pz}), $P_z$
may assume negative values as well, so that one may conclude that
geodesic particles moving along a decreasing $z$ coordinate, and
increasing $t$ coordinate, (or for that matter, an increasing $z$
coordinate and decreasing $t$, taking into account the
cosmological interpretation of Section \ref{sec:cosmo}) cannot
have come in from the exterior region.

Equation (\ref{gen-geod}) may be rewritten as
\begin{equation}
Q^2=t^2\left(\frac{V^2}{1-V^2}-\frac{P_z^2}{g_{zz}}\right) \,.
   \label{gen-geod-Q}
\end{equation}
If $\theta=\pi/2$ and $dz=0$, then the constant reduces to
\begin{equation}
Q^2=t^2\left(\frac{V^2}{1-V^2}\right) \,.
    \label{Q-zconst}
\end{equation}
For a point particle with an initial velocity $V_0<1$ and initial
time $t=T$, then $Q^2=T^2V_0^2/(1-V_0^2)$. If the particle is
initially at rest then $Q=0$.

\section{Geodesics}\label{Sec:geodesics}

An advantage of analyzing the interior region, not as a
continuation of the exterior region, but as a manifold on its own,
is a verification of the great difference existing between the
geodesics of both regions. If one treats the interior solution as
a cosmological solution, one may verify which type of universe one
is dealing with, or which geodesics are analogous with those
existing in our universe.

Consider the geodesic equation given by
\begin{equation}
\frac{d^2x^\mu}{d\lambda^2}+\Gamma^{\mu}{}_{\alpha\beta}\,
\frac{dx^\alpha}{d\lambda}\frac{dx^\beta}{d\lambda}=0   \,,
    \label{geod-eq}
\end{equation}
where $\lambda$ is an affine parameter defined along the geodesic.
It is a simple matter of exercising some index gymnastics to
verify the equivalence of the geodesic equation and the
Euler-Lagrange equations (\ref{Euler-Lagrange}).

Now, the geodesic equation, Eq. (\ref{geod-eq}), for the metric
(\ref{intmetric}) may be written in the following form
\begin{eqnarray}
&&\ddot{t}-\frac{\xi(2\xi/t-1)}{t^2}\,\dot{z}^2+\frac{m}{t^2(2\xi/t-1)}
\,\dot{t}^2+t(2\xi/t-1)\,\dot{\theta}^2
     \nonumber    \\
&&\hspace{2.5cm}+t\,\sin^2\theta \,(2\xi/t-1)\,\dot{\phi}^2=0 \,,    \\
&&\ddot{z}-\frac{2\xi}{t^2(2\xi/t-1)}
\,\dot{z}\,\dot{t}=0   \,,  \\
&&\ddot{\theta}+\frac{2}{t}\,\dot{t}\,\dot{\theta}-\sin\theta
\,\cos\theta \,\dot{\phi}^2=0  \,,  \\
&&\ddot{\phi}+\frac{2}{t}\,\dot{t}\,\dot{\phi}-2\cot\theta \,
\,\dot{\theta} \,\dot{\phi}=0   \,.
\end{eqnarray}

Considering the particular case of $\theta=\pi/2$, and using the
conserved quantities, the three primary integrals are given by
\begin{eqnarray}
&&\dot{z}=\frac{P_z}{2\xi/t-1}  \,,  \\
&&\dot{\phi}=\frac{Q}{t^2}   \,,  \\
&&\frac{\dot{t}^2}{2\xi/t-1}-\left(\frac{2\xi}{t}-1\right)\,\dot{z}^2
-t^2\,\dot{\phi}^2 =k\,,
\end{eqnarray}
which are identical to Eqs. (\ref{P})-(\ref{line1}). (See Ref.
\cite{Kiselev} for an interesting analysis of radial geodesics
confined under the Schwarzschild horizon.) We shall next analyze
null and timelike geodesics in some detail, and finally summarize
the main results in Tables \ref{t1} and \ref{t2}, respectively.

\subsection{Null geodesics}

Equation (\ref{line1}), for null geodesics, reduces to
\begin{equation}
P_z^2=\dot{t}^2-\left(\frac{Q^2}{t^2}
\right)\,\left(\frac{2\xi}{t}-1 \right)   \,. \label{nullline}
\end{equation}
Consider null geodesics along the $z-$direction, i.e., with
$d\theta=d\phi=0$, so that we simply have
\begin{equation}
P_z=\frac{dt}{d\lambda} \,,
\end{equation}
where $\lambda$ is an affine parameter defined along the geodesic.

\begin{figure}[h]
\centering
  \includegraphics[width=2.6in]{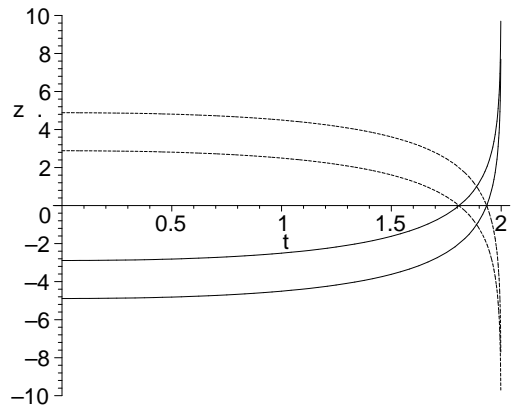}
  \caption{Plot of null geodesics, along the $z-$direction, i.e.,
  with $d\theta=d\phi=0$. See the text for details.}
  \label{Fig:null-z}
\end{figure}

For this case, the line element reduces to
\begin{equation}
ds^2=-\left(\frac{2\xi}{t}-1\right)^{-1}\,dt^2+\left(\frac{2\xi}{t}-1\right)\,dz^2
\,.
\end{equation}
Considering null geodesics, $ds^2=0$, i.e, $dt=\pm (2\xi/t-1)dz$,
we have as solution
\begin{equation}
z=\mp \left[t + 2\xi\ln\left(1-\frac{t}{2\xi}\right)\right]+C  \,,
     \label{z-null}
\end{equation}
where $C$ is a constant of integration. Equation (\ref{z-null}) is
represented in Fig. \ref{Fig:null-z}.

Note that traditionally the solution with $dt<0$ corresponds to a
black hole solution, either with an increasing or decreasing $z$
coordinate, i.e., $dz>0$ or $dz<0$, respectively. A white hole
solution corresponds to $dt>0$, either with $dz>0$ or $dz<0$. We
also emphasize the importance of analyzing the interior solution
separately, as in the literature the radial coordinate $r$
(considered in the Schwarzschild exterior) is generally considered
as a coordinate that measures distances, in the interior. It is
usually treated as a temporal coordinate to note that $r$
decreases (increases) for an observer in a black hole (white
hole).

It is also of interest to study the case of $\theta=\pi/2$ and
$dz=0$. Note that these are not circular orbits, as the $z$
coordinate can no longer be considered as a radial coordinate.
Equation (\ref{Pz}) provides $P_z=0$, and thus Eq.
(\ref{nullline}) may be rewritten as
\begin{equation}
\dot{t}^2=\left(\frac{Q^2}{t^2} \right)\left(\frac{2\xi}{t}-1
\right)   \,.
\end{equation}
The line element, for this particular case takes the form
\begin{equation}
ds^2=-\left(\frac{2\xi}{t}-1\right)^{-1}\,dt^2+t^2\,d\phi^2 \,.
\end{equation}
The null geodesic, $ds^2=0$, provides $d\phi/dt=\pm
1/\sqrt{t(2\xi-t)}$, which has the following solution
\begin{equation}
\phi=\arcsin\left(\frac{t}{\xi}-1\right)+\frac{\pi}{2}  \,,
\end{equation}
or $t=\xi[1+\sin(\phi-\pi/2)]$.
One may also obtain the equivalent solution, given by
\begin{equation}
\phi(t)=\arctan\left(\frac{t-\xi}{\sqrt{t(2\xi-t)}}\right)
+\frac{\pi}{2}  \,.
\end{equation}
The constant of integration has been chosen to provide $\phi=0$
for $t=0$. Note that for $t=2\xi$, then $\phi=\pi$. For this case
one verifies that a photon only traverses half-way around this
particular universe.


\begin{table}[h]
\begin{center}
\begin{tabular}[c]{|l|l|}
\hline
               &    \\
               & \quad $\dot{t}=\sqrt{P_z^2+\left(Q^2/t^2
\right)\,\left(2\xi/t-1 \right)}$ \; \quad \\
               & \quad $\dot{z}=\left(2\xi/t-1 \right)^{-1}\,P_z$  \\
               & \quad $\dot{\phi}=Q/t^2$   \\
               &     \\
\hline
               &     \\
\quad $dz=0$\;\; & \quad $\dot{t}=\sqrt{\left(Q^2/t^2
               \right)\,\left(2\xi/t-1 \right)}$   \\
               & \quad $\dot{\phi}=Q/t^2$   \\
               &     \\
\hline
               &      \\
\quad $d\phi=0$ & \quad $\dot{t}=P_z$   \\
               & \quad $\dot{z}=\left(2\xi/t-1 \right)^{-1}\,P_z$   \\
               &      \\
 \hline
\end{tabular}
\caption{Summary of the equations of motion for null geodesics,
considering the particular case of $\phi=\pi/2$. See the text for
details.}\label{t1}
\end{center}
\end{table}


\subsection{Timelike geodesics}

Equation (\ref{line1}), for timelike geodesics, takes the form
\begin{equation}
P_z^2=\dot{t}^2-\left(\frac{Q^2}{t^2}+1
\right)\,\left(\frac{2\xi}{t}-1 \right)   \,. \label{timeline}
\end{equation}
From the conserved quantities one may determine various
expressions relating the time coordinate and the proper time. For
instance, Eq. (\ref{timeline}) may be expressed in the following
form
\begin{equation}
d\tau=\pm
\left[P_z^2+\left(\frac{Q^2}{t^2}+1\right)\left(\frac{2\xi}{t}-1\right)
\right]^{-1/2}\;dt\,.  \label{propertime}
\end{equation}
Substituting Eq. (\ref{gen-geod}) in the above expression provides
\begin{equation}
d\tau^2=(1-V^2)\left(\frac{2\xi}{t}-1\right)^{-1} \;dt^2\,.
   \label{prop}
\end{equation}
This is an expression valid for a generic trajectory, and one
readily verifies that the variation of proper time does not depend
explicitly on the constants $P_z$ and $Q$.

One may also deduce, from Eq. (\ref{Pz}), a relationship between
the variation of proper time and the spatial coordinate, namely,
$\Delta \tau=(g_{zz}/P_z)\Delta z$. Consider the specific case of
$P_z=1$, so that $\Delta \tau=g_{zz}\Delta z$, and fixing $\Delta
z$, note that variations in proper time tend to infinity as $t
\rightarrow 0$. This is another interesting example, as viewed
from the interior, in that the test particle does not attain the
singularity in his proper time.

Taking into account the specific case of $\theta=\pi/2$ and
$d\phi=0$, which implies $Q=0$, along the direction of the $z$
coordinate, we have
\begin{eqnarray}
P_z^2=\left(\frac{dt}{d\tau} \right)^2-\left(\frac{2\xi}{t}-1
\right) \,,
     \label{Pz-Qnull}
\end{eqnarray}
from which we deduce
\begin{eqnarray}
\frac{d\tau}{dt}&=&\pm \left[\left(\frac{2\xi}{t}-1\right)+P_z^2
\right]^{-1/2}  \,.   \label{tau}
\end{eqnarray}
Taking into account the specific case of $P_z^2>1$, Eq.
(\ref{tau}) may be integrated to provide the following proper time
\begin{eqnarray}
\tau(t)&=&\pm
(P_z^2-1)^{-3/2}\Big\{-\sqrt{t(2\xi-t+P_z^2t)(P_z^2-1)}+
      \nonumber  \\
&&\hspace{-1.0cm}+\xi
\ln\left[\frac{\xi+t(P_z^2-1)}{\sqrt{P_z^2-1}}+\sqrt{t(2\xi-t+P_z^2t)}\right]\Bigg\}+C\,,
\end{eqnarray}
where $C$ is a constant of integration. If $P_z^2=1$, then the
proper time is given by
\begin{eqnarray}
\tau(t)&=&\mp \sqrt{\frac{2t}{\xi}}\;\frac{t}{3}+C\,.
\end{eqnarray}

For the particular case of $P_z^2<1$, Eq. (\ref{tau}) provides the
following solution
\begin{eqnarray}
\tau(t)&=&\pm
(1-P_z^2)^{-3/2}\Big\{-\sqrt{t(2\xi-t+P_z^2t)(1-P_z^2)}+
      \nonumber  \\
&&\hspace{-0.5cm}+\xi
\arctan\left[\frac{(1-P_z^2)t-\xi}{\sqrt{t(2\xi-t+P_z^2t)(1-P_z^2)}}\right]\Bigg\}+C\,.
\end{eqnarray}

Recall that the constant of motion may also be determined from the
initial conditions, so that substituting Eq. (\ref{Pz(V)}), with
the initial conditions $v_0$ and $T$, into Eq. (\ref{tau}), we
finally have
\begin{equation}
d\tau=\left[\left(\frac{2\xi}{T}-1\right)\left(\frac{v_0^2}{1-v_0^2}\right)
+\left(\frac{2\xi}{t}-1\right)\right]^{-1/2}\;dt   \,.
\end{equation}

The line element for $d\theta=d\phi=0$ is given by
\begin{equation}
ds^2=-\left(\frac{2\xi}{t}-1\right)^{-1}\,dt^2+\left(\frac{2\xi}{t}-1\right)\,dz^2
\,,
\end{equation}
which, taking into account Eqs. (\ref{Pz}) and (\ref{Pz-Qnull}),
takes the following form
\begin{equation}
\left(\frac{dt}{dz}\right)^{2}=\left(\frac{2\xi}{t}-1\right)^2
\left[1+\left(\frac{2\xi}{t}-1\right)/P_z^2\right] \,.
\end{equation}
In particular, for $P_z=1$, the above equation may be integrated
to yield the solution
\begin{equation}
z=\mp\frac{2(6\xi+t)}{3}\sqrt{\frac{t}{2\xi}} \pm 4\xi \;{\rm
arctanh}\left(\sqrt{\frac{t}{2\xi}}\right)\,,
\end{equation}
It may be shown that this solution is qualitatively analogous to
the plots of Fig. \ref{Fig:null-z}.

One of the most surprising results is that the trajectories of
particles at rest are geodesics, contrary to the exterior where
particles at rest are necessarily accelerated. As $P_z$ is a
conserved quantity, a particle at rest, $z={\rm const}$, will
always remain at rest. Despite the fact of the presence of strong
gravitational fields in the interior of a black hole, test
geodesic particles at rest relatively to the coordinate system may
exist, which is due to the non-static character of the interior
geometry. For an alternative approach, consider
$dz=d\theta=d\phi=0$. In this case, from $d\tau=\pm
(2\xi/t-1)^{-1/2}\,dt$, we have the following solution
\begin{equation}
\tau= \pm \sqrt{t(2\xi-t)} \mp
\xi\arctan\left[\frac{t-\xi}{\sqrt{t(2\xi-t)}}\right] +C\,.
    \label{propertime2}
\end{equation}
This solution was briefly considered in subsection
\ref{sec:cosmo}. The constant may be chosen by considering that
for $t=0$ we have $\tau=0$. For the maximum coordinate time
variation, $\Delta t=2\xi$, the corresponding proper time
variation is $\Delta \tau=\xi \pi$. This is precisely the lifetime
for the of existence of geodesic particles inside the black hole
(white hole), i.e., these test particles exist for a finite proper
time, $\xi \pi$. One verifies that Eq. (\ref{propertime2}) differs
radically from its exterior counter-part. In the exterior region
the proper time interval is inferior to the coordinate time
interval, and is interpreted as the time interval of an observer
located sufficiently far from the event horizon. A fundamental
issue is that in the exterior region, the time coordinate is
physically meaningful, as it corresponds to the proper time
measured by observers at an asymptotically large value of the
radial coordinate, $r$. In the interior region $d\tau=dt$ is but a
mere instantaneous coincidence.

For the particular case of timelike geodesic particles at rest
relatively to the $z$ coordinate, with $dz=0$ and $\theta=\pi/2$,
we have $P_z^2=0$. As emphasized above, the trajectory around the
$z-$axis cannot be interpreted as a circular orbit. The proper
time for this trajectory is determined from the following
expression
\begin{equation}
d\tau^2=\left[\left(\frac{Q^2}{t^2}+1\right)
\left(\frac{2\xi}{t}-1\right)\right]^{-1}\,dt^2  \,.
\end{equation}

The velocity of a particle along this timelike geodesic, i.e.,
$dz=0$ and $\theta=\pi/2$, as measured by an observer at rest,
taking into account Eq. (\ref{velocity2}), is given by
\begin{equation}
V^2= -\frac{g_{\phi\phi}}{g_{tt}}\left(\frac{d\phi}{dt}\right)^2
=\left(\frac{2\xi}{t}-1 \right)\,\frac{Q^2}{t^2 \dot{t}^2}=
\frac{Q^2}{t^2+Q^2}\,.
\end{equation}
This expression may also be obtained from Eq. (\ref{Q-zconst}).
Note that as $t \rightarrow 0$, then $V \rightarrow 1$. At
$t=2\xi$, we verify that the particle attains a finite minimum
value, given by $V^2=Q^2/(Q^2+4\xi^2)$.

For the particular case of $dz=d\phi=0$ and $\theta=\pi/2$, we
verify that the constants of motion are zero, $P_z=Q=0$, implying
that the timelike geodesic particles remain at rest. An important
conclusion is inferred from the conserved quantities for particles
at rest. As is well known, an incoming geodesic particle from the
exterior, has a conserved quantity $E$, which is interpreted as
the energy per unit mass, along its trajectory. However, this
constant of motion in the interior of the event horizon changes
its physical significance, but its numerical value remains
invariant. If $P_z=0$ is verified, this is equivalent to state
that the test particle entered from the exterior with $E=0$. Now,
the energy per unit mass is defined as $E=(1-2M/r)/(1-v_0^2)$, so
that $E=0$ corresponds to $r=2M$. This means that the particle
started off from the horizon, which is a null surface. Thus, for
the particular case of $P_z=0$, one may conclude that a geodesic
timelike particle at rest in the interior of the horizon cannot
have come in from the exterior region.


\begin{table}[h]
\begin{center}
\begin{tabular}[c]{|l|l|}
\hline
               &         \\
               & \quad $\dot{t}=\sqrt{P_z^2+\left(Q^2/t^2+1
                     \right)\,\left(2\xi/t-1 \right)}$ \hspace{0.1cm}  \\
               & \quad $\dot{z}=\left(2\xi/t-1 \right)^{-1}\,P_z$   \\
               & \quad $\dot{\phi}=Q/t^2$   \\
               &         \\
\hline
               &         \\
\quad $dz=0$ \;        & \quad
                 $\dot{t}=\sqrt{\left(Q^2/t^2+1
                 \right)\,\left(2\xi/t-1 \right)}$   \\
               & \quad $\dot{\phi}=Q/t^2$   \\
               &          \\
\hline
               &          \\
\quad $d\phi=0$      & \quad $\dot{t}=\sqrt{P_z^2+\left(2\xi/t-1 \right)}$   \\
                     & \quad $\dot{z}=\left(2\xi/t-1 \right)^{-1}\,P_z$   \\
               &          \\
 \hline
                     &            \\
\quad $dz=0$         &   \quad $\dot{t}=\sqrt{\left(2\xi/t-1 \right)}$       \\
\quad $d\phi=0$      &            \\
                     &            \\
 \hline
\end{tabular}
\caption{Summary of the equations of motion for timelike
geodesics, considering the particular case of $\phi=\pi/2$. See
the text for details.}\label{t2}
\end{center}
\end{table}


\section{Eddington-Finkelstein
coordinates}\label{Sec:EK-coordinates}

The Eddington-Finkelstein transformation is traditionally
considered a transformation that permits the analysis of
trajectories from $0<r<\infty$. However, in a general manner, the
inversion of the character of the coordinates is not manifest.
Therefore, to manifest this difference, we shall treat the
Eddington-Finkelstein transformations directly from the interior
metric (\ref{intmetric}).

For null geodesics along the $z-$direction, Eq. (\ref{z-null})
provides the following solutions
\begin{eqnarray}
z_\mp=\mp \left[t+2\xi\ln\left(1-\frac{t}{2\xi}\right)\right]+C
\,.
\end{eqnarray}
The solution with the negative sign shows that $z$ increases as
$dt<0$, and decreases as $dt>0$; from the solution with the
positive sign, one may infer that $z$ increases as $dt>0$, and
decreases as $dt<0$.

Consider now the following transformations
\begin{eqnarray}
z'&=&z_- +2\xi\ln\left(1-\frac{t}{2\xi}\right)
\quad \Rightarrow \quad z'=-t+C  \,,   \label{zminus2} \\
z''&=&z_+ -2\xi\ln\left(1-\frac{t}{2\xi}\right)
\quad \Rightarrow
\quad z''=t+C   \,. \label{zplus2}
\end{eqnarray}

In the exterior region of the event horizon, solutions for $dt<0$
are excluded, as one admits that the temporal coordinate
increases. In the interior region two distinct cases need to be
separated, namely, for $dt<0$, which traditionally is denoted a
black hole, and $dt>0$, a white hole.

Taking into account the definition $z'$, one may rewrite the
metric (\ref{intmetric}) as
\begin{equation}
ds^2=\left(\frac{2\xi}{t}-1\right)\,dz'^2+\frac{4\xi}{t}\,dz'dt
+\left(\frac{2\xi}{t}+1\right)\,dt^2+t^2\,d\Omega^2
       \label{intmetricEF}    \,,
\end{equation}
which is no longer singular at $t=2\xi$.

Now metric (\ref{intmetricEF}) may be simplified by introducing a
null coordinate, denoted the advanced time parameter in analogy
with the exterior solution
\begin{eqnarray}
v'=z'+t = z_-+2\xi\,\ln\left(1-\frac{t}{2\xi}\right)+t \,,
     \label{def-v}
\end{eqnarray}
so that the metric (\ref{intmetric}) takes the form
\begin{equation}
ds^2=\left(\frac{2\xi}{t}-1\right)\,dv'^2+2\,dtdv'+t^2\,d\Omega^2
       \label{intmetricEF2}    \,.
\end{equation}
This is the line element of Eddington-Finkelstein for the advanced
time parameter, which is regular at the instant $t=2\xi$.

\begin{figure}[h]
\centering
  \includegraphics[width=2.8in]{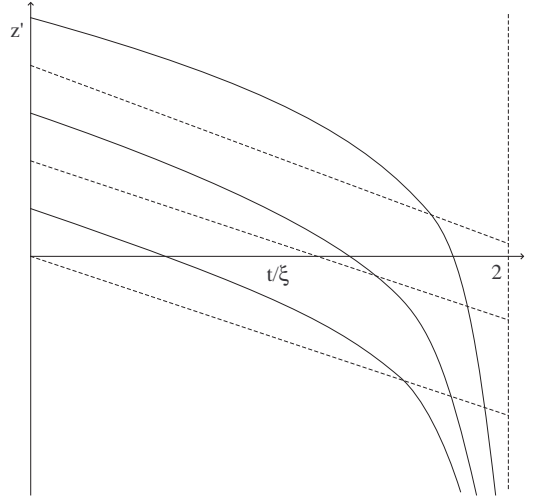}
  \caption{Eddington-Finkelstein diagram for the advanced temporal
parameter. A black hole solution corresponds to $dt<0$, and
consequently $dz'>0$; and analogously, a white hole solution
corresponds to $dt>0$ and $dz'<0$.}
  \label{Fig:Edding-Finkel}
\end{figure}

Analyzing the specific case of $ds^2=d\theta=d\phi=0$, the metric
(\ref{intmetricEF2}) provides the following solutions
\begin{equation}
dv'=0, \qquad {\rm or} \qquad
\left(\frac{2\xi}{t}-1\right)dv'=-2\,dt
   \,.
\end{equation}
Recalling that $dv'=dz'+dt$, the above cases with
\begin{eqnarray}
dz'&=&-dt   \,, \\
\frac{dt}{dz'}&=&-\frac{2\xi-t}{2\xi+t} \,,
\end{eqnarray}
have the following solutions
\begin{eqnarray}
z'&=&-t + C   \,, \\
z'&=&t+4\xi \ln\left(1-\frac{t}{2\xi}\right)+C \,.
\end{eqnarray}
These are plotted in Fig. \ref{Fig:Edding-Finkel}, for different
values of the constant $C$. Note that both solutions obey
$dz'/dt<0$. A black hole solution corresponds to $dt<0$, and
consequently $dz'>0$; and analogously, a white hole solution
corresponds to $dt>0$ and $dz'<0$.

Applying an analogous procedure for the retarded temporal
parameter, $w''$, constructed from $z''$,
\begin{equation}
w'=z''-t \Rightarrow
w'=z_+-2\xi\,\ln\left(1-\frac{t}{2\xi}\right)-t \,,
     \label{def-w}
\end{equation}
and consequently
\begin{equation}
ds^2=\left(\frac{2\xi}{t}-1\right)\,dw'^2-2\,dtdw'+t^2\,d\Omega^2
       \label{intmetricEF3}    \,.
\end{equation}
As is manifest from the line elements (\ref{intmetricEF2}) and
(\ref{intmetricEF3}), the metric coefficients are regular at
$t=2\xi$.

\begin{figure}[h]
\centering
  \includegraphics[width=2.8in]{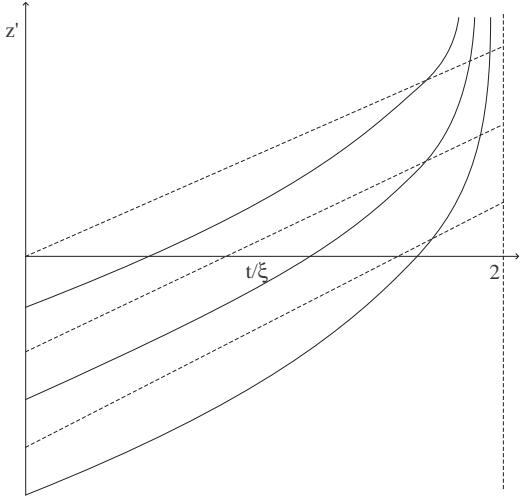}
  \caption{Eddington-Finkelstein diagram for the retarded temporal
parameter. A black hole solution corresponds to $dt<0$, and
consequently $dz'<0$; and analogously, a white hole solution
corresponds to $dt>0$ and $dz'>0$.}
  \label{Fig:Edding-Finkel2}
\end{figure}

For the case $ds^2=d\theta=d\phi=0$, the metric
(\ref{intmetricEF3}) provides the following relationships
\begin{eqnarray}
dz''&=&dt   \,, \\
\frac{dt}{dz''}&=&\frac{2\xi-t}{2\xi+t} \,,
\end{eqnarray}
with the respective solutions
\begin{eqnarray}
z''&=&t + C   \,, \\
z''&=&-t-4\xi \ln\left(1-\frac{t}{2\xi}\right) \,.
\end{eqnarray}
These are plotted in Fig. \ref{Fig:Edding-Finkel2}, for different
values of the constant $C$. Both solutions obey $dz''/dt>0$, with
$dt<0$ and $dz''<0$ corresponding to a black hole solution; and
$dt>0$ and $dz''>0$ to a white hole solution, respectively.

\section{Kruskal coordinates}\label{Sec:Kruskal-coordinates}

Consider the difference obtained from Eqs. (\ref{def-v}) and
(\ref{def-w}), given by
\begin{equation}
v'-w'=4\xi \ln\left(1-\frac{t}{2\xi}\right)+2t  \,,
\end{equation}
from which one may obtain the following equalities
\begin{eqnarray}
1-\frac{t}{2\xi}&=&\exp\left(\frac{v'-w'}{4\xi}\right)
\exp\left(-\frac{t}{2\xi}\right)
   \label{def1}    \,, \\
dt&=&-(dv'-dw')\frac{2\xi-t}{2t}\,.
\end{eqnarray}
Substituting these expressions in Eq. (\ref{intmetricEF2}), one
obtains
\begin{equation}
ds^2=\left(\frac{2\xi}{t}-1\right)dw'dv'+t^2(d\theta^2+\sin^2\theta
\, d\phi^2) \,.    \label{intmetricEF4}
\end{equation}

Now, introducing the Kruskal coordinates for the region $t<2\xi$
($r<2M$), i.e.,
\begin{eqnarray}
w''&=&\exp\left(-\frac{w'}{4\xi}\right)  \,, \label{def-w''}\\
v''&=&\exp\left(\frac{v'}{4\xi}\right)\,.    \label{def-v''}
\end{eqnarray}
Substituting these expressions in Eq. (\ref{def1}), we finally
have
\begin{equation}
w''
v''=\left(1-\frac{t}{2\xi}\right)\exp\left(\frac{t}{2\xi}\right)
\,.
\end{equation}

Equations (\ref{def-w''})-(\ref{def-v''}) may be rewritten as
\begin{eqnarray}
dw'&=&-\frac{dw''}{w''}\,4\xi   \,, \label{def-w''2}\\
dv'&=& \frac{dv''}{v''}\,4\xi \,,   \label{def-v''2}
\end{eqnarray}
which substituting into metric (\ref{intmetricEF4}), we have the
following
\begin{equation}
ds^2=-\frac{32\xi^3}{t}
\exp\left(-\frac{t}{2\xi}\right)\;dw''dv''+t^2(d\theta^2+\sin^2\theta
\, d\phi^2)  \,.
\end{equation}

It is still possible to introduce the following transformations
\begin{eqnarray}
t_*&=&\frac{1}{2}(v''+w'')   \,, \label{def-t'}\\
r_*&=&\frac{1}{2}(v''-w'')   \,,   \label{def-r'}
\end{eqnarray}
so that we have $dt_*^{2}-dr_*^{2}=dv''\,dw''$. The line element
finally assumes the form
\begin{equation}
ds^2=\frac{32\xi^3}{t}
\exp\left(-\frac{t}{2\xi}\right)\;\left(-dt_*^{2}+dr_*^{2}\right)+t^2\,d\Omega^2
\,.
\end{equation}

\begin{figure}[h]
\centering
  \includegraphics[width=3.2in]{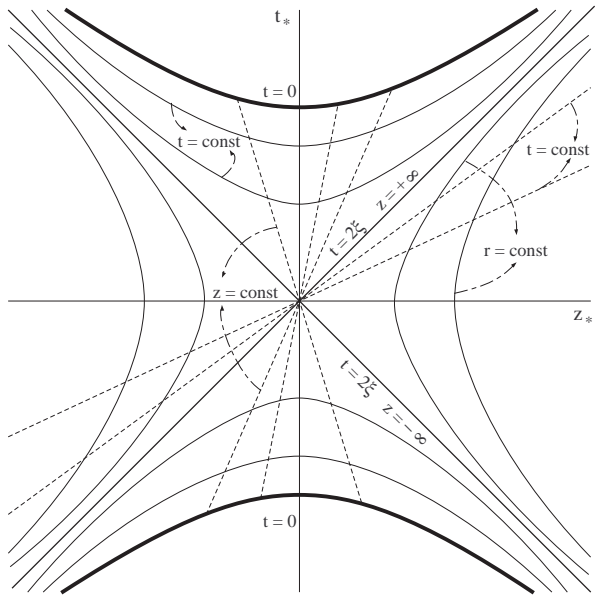}
  \caption{Kruskal diagram for the interior region of the
  Schwarzschild spacetime. We have added some details in the
  exterior region for comparison. See the text for details.}
  \label{fig:diagram}
\end{figure}

The new coordinates may be rewritten as
\begin{eqnarray}
t_*&=&\frac{1}{2}\left[\exp\left(\frac{v'}{4\xi}\right)
+\exp\left(-\frac{w'}{4\xi}\right) \right]
      \nonumber  \\
&=&\left(1-\frac{t}{2\xi}\right)^{1/2}\exp\left(\frac{t}{4\xi}\right)\,\cosh\left(\frac{z}{4\xi}\right)
                      \label{def-t'2}\\
r_*&=&\frac{1}{2}\left[\exp\left(\frac{v'}{4\xi}\right)
-\exp\left(-\frac{w'}{4\xi}\right) \right]
      \nonumber \\
&=&\left(1-\frac{t}{2\xi}\right)^{1/2}\exp\left(\frac{t}{4\xi}\right)\,\sinh\left(\frac{z}{4\xi}\right)\,.
  \label{def-r'2}
\end{eqnarray}
These expressions may be written
\begin{equation}
t_*^{2}-r_*^{2}=\left(1-\frac{t}{2\xi}\right)\,\exp\left(\frac{t}{2\xi}\right)
\,,
\end{equation}
which is the equation for a hyperbole, and may also be expressed
as
\begin{equation}
\frac{r_*}{t_*}=\tanh\left(\frac{z}{4\xi}\right) \,,
\end{equation}
which represent straight lines with $z={\rm const}$. See Fig.
\ref{fig:diagram}.

The singularity at $t=0$, written in terms of the new coordinates,
is given by
\begin{equation}
t_*=\pm \sqrt{r_*^2+1}  \,.
\end{equation}
For $r_*=0$, we have $t_*=1$. For $t=2\xi$, we have $t_*=\pm r_*$,
i.e., $\tanh(z/4\xi)=\pm 1$, which implies $z \rightarrow \pm
\infty$. These relationships may be visualized in Fig.
\ref{fig:diagram}. We have also added the exterior region, for
comparison purposes (see, for instance, Ref. \cite{Wald}).

With this analysis at hand, one may consider the following motion
of a timelike test particle, as viewed from an interior observer.
The test particle starts its movement at the event $E_1$, arriving
at the surface of $t=2\xi$ and $z=-\infty$, at event $E_2$. After
an excursion in the exterior region, the test particle re-enters
into the interior region at event $E_3$, corresponding to $t=2\xi$
and $z=+\infty$, and finally ends up in the spacelike singularity
at $t=0$, at event $E_4$. Note an extremely curious feature of
this movement, as viewed from an interior observer. The test
particle exits the interior region, at the moment of complete
contraction along the negative end of the $z$ direction, to
reappear instantaneously at $t=2\xi$, at the positive side of the
$z$ axis. According to the point of view of the interior observer,
no time has elapsed during the test particle's excursion in the
exterior region. This analysis is analogous to the one outlined in
Ref. \cite{Brehme}. Another curious feature, relatively to the
interior observer is also worth mentioning: All infalling null or
timelike particles enter into the interior at different places
$z=\pm \infty$, but {\it simultaneously} at $t=2\xi$.

\begin{figure}[h]
\centering
  \includegraphics[width=3.2in]{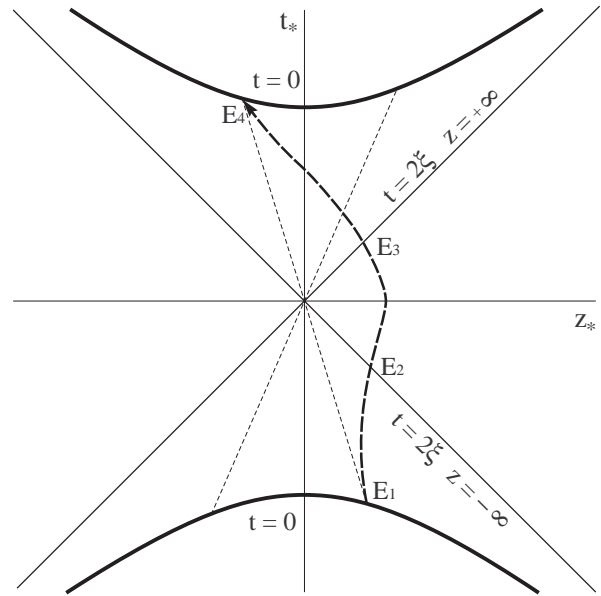}
  \caption{A hypothetical test particle starts its movement at the
  event $E_1$, i.e., at $t=0$, and arrives at $t=2\xi$ and $z=-\infty$, at event $E_2$.
  It re-enters into the interior region at event $E_3$, corresponding
to $t=2\xi$ and $z=+\infty$, ending up in the spacelike
singularity at $t=0$, at event $E_4$. As viewed from an interior
observer the test particle exits the interior region, at $t=2\xi$
and $z=-\infty$, to reappear instantaneously at $t=2\xi$, at the
positive side of the $z$ axis. According to the point of view of
the interior observer, no time has elapsed during the test
particle's excursion in the exterior region.}
  \label{fig:diagram2}
\end{figure}

\section{Summary and discussion}\label{Sec:conclusion}

The Schwarzschild solution has played a fundamental conceptual
role in general relativity, and beyond, for instance, regarding
event horizons, spacetime singularities and aspects of quantum
field theory in curved spacetimes.
In this work, we have provided a brief pedagogical review and
further analyzed the interior Schwarzschild solution. Firstly, by
deducing the interior metric by considering time-dependent metric
fields, we have analyze the interior region, without the
prejudices inherited from the exterior region. With this geometry
at hand, we have payed close attention to several respective
cosmological interpretations, and addressed some of the
difficulties associated to spacetimes singularities. Secondly, we
have deduced the conserved quantities of null and timelike
geodesics, and discussed several particular cases in some detail.
Finally, we examined the Eddington-Finkelstein and Kruskal
coordinates directly from the interior solution.

A black hole is believed to have formed from the gravitational
collapse of a massive body. However, events occurring in the
interior of the event horizon are not observable for an exterior
observer, and one may argue that relatively to the latter, black
holes are not relevant physical objects \cite{Narlikar}. Although
the event horizon exists for exterior observers, all events in the
range $r>2M$ are accessible to the interior observers.
If one looks at the interior geometry as a continuation of the
exterior static solution, one comes across some extremely
interesting conceptual difficulties, that question the very
concept of a black hole. For instance, while for the exterior
observer, infalling particles end up at a central singularity at
$r=0$, from the interior point of view, the proper distance along
the $z-$direction increases, showing the existence of a cigar-like
singularity. The latter singularity is a spacelike hypersurface,
and the test particles are not directed towards a privileged
point, however, in order to not violate causality they are
directed along a temporal direction from $t=2\xi$ to $t=0$. A
curious behavior relatively to an interior observer is also
verified, as all infalling particles crossing the event horizon,
occur simultaneously at $t=2\xi$.
In this context, the Eddington-Finkelstein and Kruskal
transformations do indeed solve the coordinate singularity at
$r=2M$, but do not solve the problems associated with the
inversion of the $r$ and $t$ coordinates. Assuming that $r$ is a
temporal coordinate for $r<2M$, also signifies giving it a
determined direction and duration, i.e., the black hole, or for
that matter a white hole, possesses a finite coordinate temporal
duration. However, the exterior geometry is static, and once
created does not disappear.

An interesting feature relatively to the interior geometry is the
issue of proper distances. The proper distance between two
particles at rest separated by a constant $\Delta z$, decreases
along the $z-$direction as coordinate time flows from $t=0$ to
$t=2\xi$, and increases as coordinate time flows backwards from
$t=2\xi$ to $t=0$. In counterpart, the proper distance between two
simultaneous events along a spatial trajectory with $dz=0$ and
$\theta=\pi/2$, increases as $t$ varies from $t=0$ to $t=2\xi$,
and decreases when the temporal coordinate runs backwards from
$t=2\xi$ to $t=0$. Another surprising result, considering the
interior point of view, is that the trajectories of particles at
rest are geodesics, contrary to the exterior where particles at
rest are necessarily accelerated. This fact is due to the
non-static character of the interior geometry.

In this work, we have addressed some conceptual difficulties
related to the notion of black holes. The solutions that do away
with the interior singularity and the event horizon
\cite{Mazur,gravastars,Dymnikova,darkstars}, although interesting
in themselves, sweep the inherent conceptual difficulties of black
holes under the rug. In concluding, we note that the interior
structure of realistic black holes have not been satisfactorily
determined, and are still open to considerable debate.



\end{document}